\documentclass[aps,prd,11pt,showpacs,showkeys,preprintnumbers,superscriptaddress,nobibnotes,floatfix,longbibliography]{revtex4-1}

\pdfoutput=1
\usepackage{amsmath,amsfonts,amssymb,mathrsfs,graphicx,color,longtable}
\usepackage{hyperref}

\usepackage{mathrsfs}
\usepackage{hyperref}
\usepackage{graphicx}
\usepackage{amsfonts,amsmath,amssymb,bm}
\usepackage{array}
\usepackage{color}
\usepackage{enumitem}
\usepackage[explicit]{titlesec}
\usepackage[utf8]{inputenc}
\usepackage{multirow}
\usepackage{lineno}
\usepackage{ulem}
\usepackage[caption=false]{subfig}
\usepackage{upgreek}

\usepackage{threeparttable}

\def\tabnotefont{\fontsize{9}{10}\selectfont}%

\usepackage{lineno}

\begin{document}

\title{Quantitative U/Th deposition and cleanliness control strategies in the JUNO site air}

\author{Jie Zhao}
\email{zhaojie@ihep.ac.cn}
\affiliation{Institute of High Energy Physics, Chinese Academy of Sciences, Beijing, China}

\author{Chenyang Cui}
\affiliation{Institute of High Energy Physics, Chinese Academy of Sciences, Beijing, China}
\affiliation{University of Chinese Academy of Sciences, Beijing, China}

\author{Yongpeng Zhang}
\affiliation{Institute of High Energy Physics, Chinese Academy of Sciences, Beijing, China}

\author{Gaosong Li}
\affiliation{Institute of High Energy Physics, Chinese Academy of Sciences, Beijing, China}

\author{Nan Wang}
\affiliation{Institute of High Energy Physics, Chinese Academy of Sciences, Beijing, China}

\author{Monica Sisti}
\affiliation{INFN Milano Bicocca and Universit\`{a} di Milano-Bicocca, Milano, Italy}

\begin{abstract}
The Jiangmen Underground Neutrino Observatory (JUNO) employs a 20~kt liquid scintillator (LS) detector located 700~m underground. To meet its physics objectives, the LS must achieve an ultra-low $^{238}$U/$^{232}$Th content of 10$^{-17}$~g/g. Given that airborne dust exhibits radioactivity about 12 orders of magnitude higher, exceptional cleanliness is essential during on-site installation. The total permissible dust mass in the 20~kt LS is only about 8~mg. To attain this, the acrylic vessel interior must comply with class~1,000 cleanliness. Pre-filling water spray cleaning improves cleanliness by roughly two orders of magnitude, requiring the overall environment to be maintained between class~10,000 and 100,000. At JUNO, a cleanroom management system has been implemented across the 120,000~m$^3$ underground experimental hall. Since May 2022, continuous laser particle monitoring has consistently achieved an average cleanliness class of 74,000. Furthermore, we developed a method to directly measure $^{238}$U/$^{232}$Th deposition rates on detector surfaces. Using ICP-MS, sensitivity reaches sub-ppt levels ($<$10$^{-12}$~g/g), enabling effective cleanliness control and assessment of external contamination during detector construction.
\end{abstract}

\maketitle

\section{Introduction}
Dust in the air typically contains radioactive $^{238}$U/$^{232}$Th at the level of ppb-ppm. Therefore, for ultra-low background detection technology and low-background experiments, there are strict requirements for environmental cleanliness. The underground rock at the Jiangmen Underground Neutrino Observatory (JUNO) site contains 9.6~ppm of $^{238}$U and 26.6~ppm of $^{232}$Th~\cite{JUNO:2021vlw}. The walls of the experimental hall have only been sandblasted and not specially protected, allowing rock powder to disperse into the air and settle onto the surface of the detector during construction activities and ventilation. 

The main target medium of the JUNO experiment is a 20~kt liquid scintillator (LS) contained in a 35.4-meter-diameter acrylic sphere. To meet all JUNO physics requirements, the $^{238}$U/$^{232}$Th content in the LS is required to reach a level of 10$^{-17}$~g/g~\cite{JUNO:2021vlw}, which is about 12 orders of magnitude lower than the $^{238}$U/$^{232}$Th content in the rock. The dust present on the inner surface of the acrylic vessel and within the internal environment were in direct contact with the LS during the filling, thus we imposed extremely high cleanliness requirements on the acrylic sphere and its surroundings. Additionally, if too much dust accumulates on the detector components outside the acrylic sphere, such as Photomultiplier Tubes (PMTs), the Stainless Steel (SS) structure, and Ultrapure Water (UPW), the gamma radiation resulting from $^{238}$U/$^{232}$Th decay in the dust can penetrate the water layer (see Fig.~\ref{fig:detector}) and contribute to the background in the LS. This necessitates stringent cleanliness requirements not only for the acrylic vessel containing the LS but also for the entire installation environment in the hall. 

Furthermore, in an environment requiring a certain level of cleanliness, it is necessary to quantitatively measure how much natural radioactive $^{238}$U/$^{232}$Th will settle on the detector. Inductively Coupled Plasma Mass Spectrometry (ICP-MS) has extremely high sensitivity for measuring $^{238}$U/$^{232}$Th, allowing for the design of experiments to quantitatively measure the actual $^{238}$U/$^{232}$Th deposition rate under known levels of environmental cleanliness. 

This work will provide a detailed overview of the cleanliness control strategies implemented at the JUNO site and the current level of cleanliness achieved. 
The structure of the article is as follows: Section~\ref{sec2} presents the conceptual design of the JUNO detector. Section~\ref{sec3} explains how the radiopurity requirements for the JUNO LS were translated into cleanliness standards for the air in direct contact with the LS inside the acrylic vessel, and how these standards were extended to the entire experimental hall environment.
Building on these criteria and considering actual installation conditions, Section~\ref{sec4} describes the cleanliness control strategy developed for the installation process and evaluates its effectiveness through long-term monitoring of particulate levels in the hall. Section~\ref{sec5} presents a precise experiment conducted during the installation phase to measure the U/Th deposition on surfaces of various shapes and orientations over a defined period. This experiment enabled the quantification of expected U/Th accumulation on detector surfaces throughout the installation. Finally, based on measured U/Th deposition rates and simulations of natural radioactivity at different detector material locations, Section~\ref{sec6} estimates the background contribution in the LS from U/Th deposited on detector surfaces during installation, as well as its impact on radon contamination levels in the external UPW region. Section~\ref{sec7} is the summary of this work.

\section{Overview of the JUNO detector}
\label{sec2}

\begin{figure}[!h]
    \centering
    \includegraphics[width=0.7\textwidth]{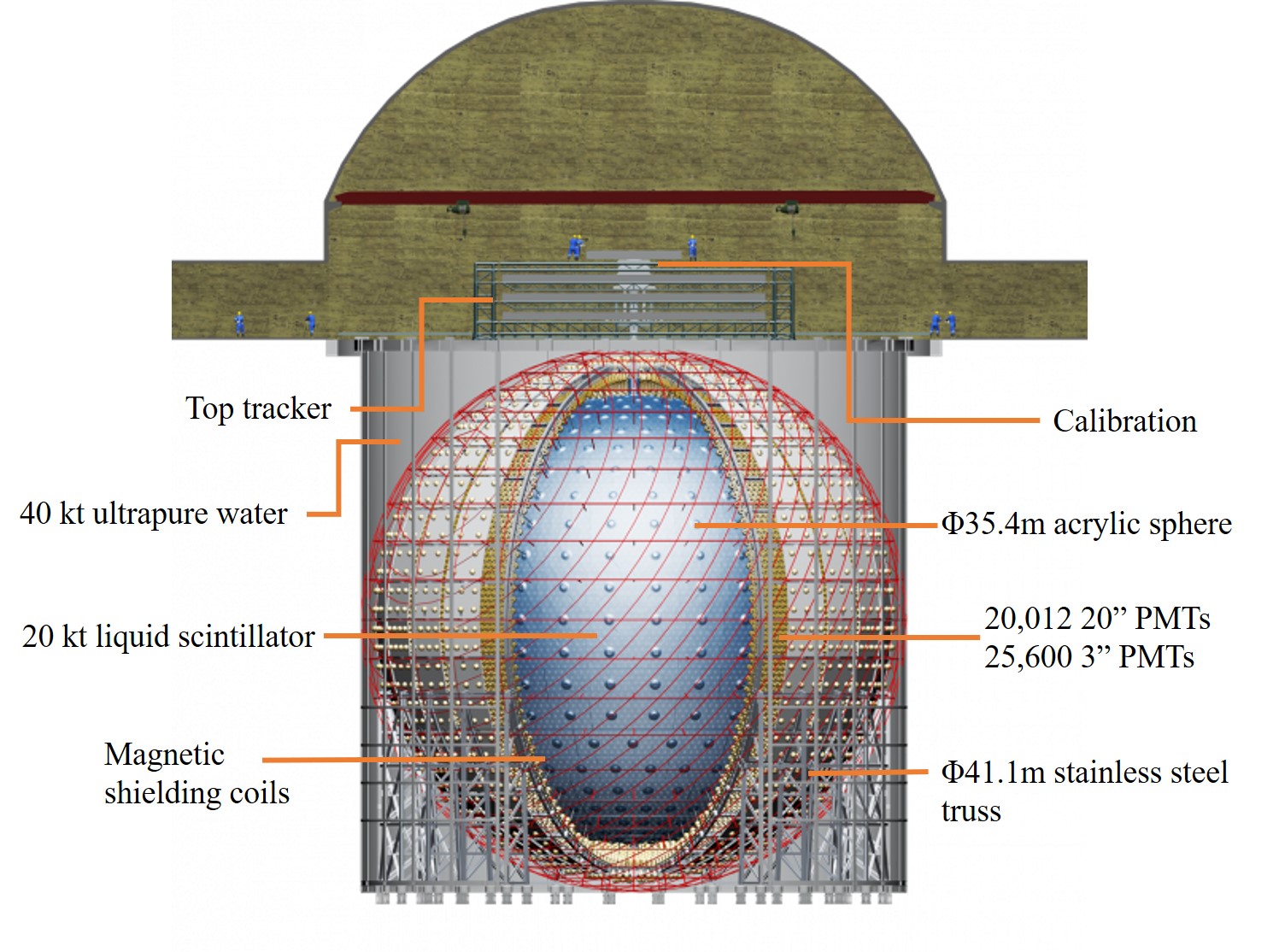}
    \caption{Conceptual diagram of the JUNO detector.}
    \label{fig:detector}
\end{figure}

The conceptual design of the JUNO detector is illustrated in Fig.~\ref{fig:detector}. A 20~kt LS, serving as the target for neutrino detection, is contained within a 35.4-meter-diameter acrylic sphere. The entire acrylic sphere is supported by a 41.1-meter-diameter SS truss, which is held in place by SS support rods. A total of 17,612 20-inch and 25,600 3-inch PMTs are installed on the SS truss, designed to detect the faint light signals generated by neutrino interactions with the LS. The entire spherical structure is housed within a cylindrical cavity measuring 43.5~m in diameter and 44~m in height. This cavity is filled with 40~kt of UPW, which, in combination with the top-mounted track detector, serves to tag cosmic rays and shield against environmental background. Additionally, magnetic shielding coils are installed on the SS truss to mitigate the interference of the Earth's magnetic field on the PMTs. A calibration room is positioned at the top of the spherical detector, approximately 9 meters above the surface of the acrylic sphere. Due to the lower density of the LS compared to water, the liquid level inside the acrylic sphere is maintained 4.5~m higher than the external water level to ensure pressure equilibrium. Consequently, a 9-meter chimney extends upward from the top of the acrylic sphere. To account for light blocking, mechanical strength, and background effects, the lower 1~m section of the chimney is made of acrylic, while the remaining portion is constructed of SS. The chimney's upper end extends into the top calibration room, allowing calibration sources to be introduced into the detector's interior through the chimney for calibration purposes.

\section{Cleanliness requirement}
\label{sec3}
JUNO high-purity scenario requires the $^{238}$U/$^{232}$Th concentration in its 20~kt LS to be at the level of 10$^{-17}$~g/g~\cite{JUNO:2021vlw}, corresponding to a maximum of 0.2~$\upmu$g for both $^{238}$U and $^{232}$Th in the full detector volume. 

The inner surface of the acrylic sphere was protected by a film (purchased from Daio Paper Corporation~\cite{film}) during the construction process. The main material of this film is paper, with one side coated with a thin layer of water-soluble adhesive. During the final cleaning process, the adhesive dissolves, allowing the paper film attached to the surface of the organic glass to naturally fall off. Therefore, before removing the protective film, it is possible to use 3D nozzles to generate micron-level water mist, allowing the dust in the air inside the sphere to adhere to the water droplets and settle to the bottom of the acrylic sphere, thereby further improving the air quality inside the sphere. Based on usual observations, the concentration of particulate matter in urban air significantly improves immediately after rain, typically reducing by one order of magnitude. With better optimization and design at JUNO site, we expect the effect of dust reduction through water mist in the 35.4-meter-diameter acrylic sphere allows an improvement of one to two orders of magnitude in the cleanliness. If the initial cleanliness level inside the sphere is maintained within Class 10,000–100,000, it can be further improved to better than Class 1,000 after the water mist treatment. 

After the completion of the detector construction, the upper and lower chimney openings were temporarily sealed with a plate to avoid significant exchange of internal and external gases. The JUNO acrylic sphere will first be filled with UPW from the bottom, then the LS will be injected from the top to replace the UPW inside the sphere. During the water filling process, the air inside the sphere will come into direct contact with the acrylic surface and the UPW. The most conservative assumption is that all the dust in the air inside the sphere will be transferred to the LS. The dust in the air mainly comes from rock powder, with a radioactive assumption consistent with that of the rock (9.6~ppm $^{238}$U and 26.6~ppm $^{232}$Th). Therefore, it is required that the dust content in the air be less than 8~mg, dominated by $^{232}$Th concentration. According to the particle size distribution in the cleanroom, assuming that the particles are spherical and their density is similar to the rock of 2.6~g/cm$^3$, the total particle mass in a cleanroom of class~1,000 (equivalent to ISO class 6 according to "Federal Standard 209E") and size 23,216~m$^3$ is 12~mg, which is close to the requirement for the LS. During the actual filling process, the air inside the sphere will be vented from the top, and the dust in the air will not all settle on the water and acrylic surfaces. However, we conservatively require the cleanliness of the air inside the sphere to reach class~1,000 before filling it with water.

\section{Cleanliness control strategies}
\label{sec4}
\subsection{Overall environment}
In early 2022, the SS truss construction began in an installation environment that was still very poor at that time. The hall and ancillary rooms were not sealed and were connected to the outside traffic support tunnel. The rock walls inside the hall were only sandblasted, without any other special protection. The weathered rock powder on the surface of the rock walls would float into the air with the airflow, and the activities of the installation personnel would also bring dust. Air cleanliness was measured with a particle counter, and the particle concentration is estimated to be comparable to that of Class 1,000,000. There were four self-circulating filtration fans in the hall with a circulation and filtration capacity of 200,000~m$^3$/h to improve the cleanliness of the hall. All self-circulating filtration fans are equipped with a three-stage filtration system, consisting of primary, medium, and high-efficiency filters. However, the hall is not isolated from the external environment, and there are no clean protective measures for the movement of personnel and goods. As a result, even with four self-circulating filtration fans, the capacity for environmental improvement remains limited.

In order to improve the cleanliness of the hall environment, we carried out a deep cleaning of the hall in April 2022. The rock walls above the ground level of 3~m were cleaned with water to reduce dust caused by personnel activities. For areas above 3~m, water flushing was challenging on one hand, and the ducts had sponges, which could not be flushed with a large amount of water. However, residual dust in the experimental hall is expected to gradually diminish over time through continuous circulation and filtration. All surfaces above and below 3~m, such as the floor, handrails, and ducts were vacuumed and then wiped again. The walls of the cylindrical water pool, with a diameter of 43.5~m and height of 44~m, were rinsed with deionized water, and the installed SS structure was also rinsed with deionized water. 

The cargo air shower rooms at the top and bottom entrances of the hall were completed and put into operation in early May 2022. From then on, all goods and personnel entering the hall needed to pass through the air shower to remove surface dust. In the same month, fire doors were installed in the ancillary rooms around the hall, effectively achieving isolation between the hall and the outside traffic support tunnel. At the same time, we set up dressing areas at the entrances of the two cargo air shower rooms to implement a clean room management mode in the hall. Personnel entering the hall needed to change into special anti-static clean suits in this area. Plastic mats were laid on the cement floor inside the hall to facilitate cleaning and reduce dust caused by personnel movement on the cement floor.

In early July 2022, the SS truss was completed, and preparations were made to begin the assembly of the acrylic sphere. Prior to this, we carried out a second cleaning of the truss, using vacuum cleaners for dust removal and wiping with a damp cloth. At the same time, we inspected the rusting situation in certain areas of the SS truss, and treated the rust spots by removing the rust and applying polyurea protection.

The acrylic sphere has the highest cleanliness requirement. Therefore, starting from the installation of the acrylic sphere, we have implemented strict cleanroom management in the experimental hall. The main measures are as follows: 

\begin{itemize}
    \item All personnel entering the hall must wear dedicated cleanroom suits, hats, and shoe covers. 
    \item All goods and equipment entering the hall must be cleaned outside the hall, wrapped in protective film during transportation, and the film must be removed at the hall entrance and passed through the air shower in the cargo air shower room before entering the hall. 
    \item The paper outer packaging of all goods and equipment must be removed at the entrance of the cargo air shower room before entering, and only low-dust paper boxes customized for PMT package are allowed to enter. 
    \item Components waiting for installation in the hall must be stored on dedicated plastic trays isolated from the ground, and the surface of the components must be covered with anti-static clean cloth for protection. All tools or small installation components in the hall must be placed in dedicated plastic boxes and covered with clean cloth for protection. 
    \item Welding and cutting operations are not allowed in the entire hall and the surrounding traffic support tunnel. 
\end{itemize}

In terms of management, the overall communication and coordination are the responsibility of the head of the background task force, the heads of the various subsystems are responsible for review and implementation, and the quality inspection is carried out by a gatekeeper. 

\begin{figure*}[h]
    \centering
    \includegraphics[width=0.9\textwidth]{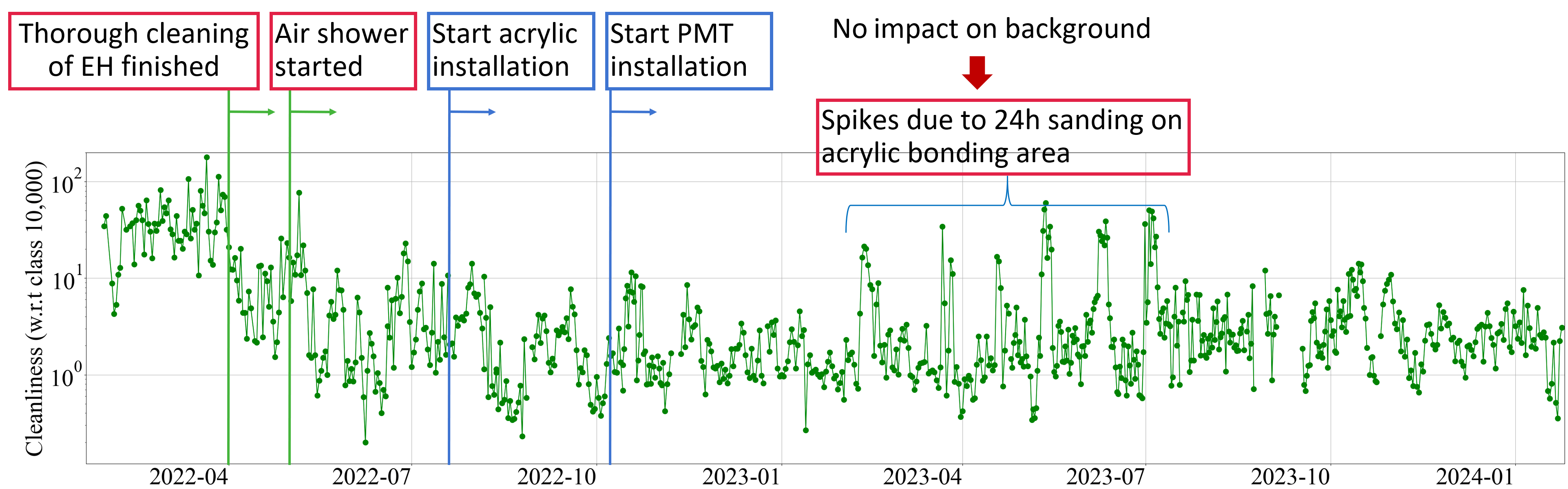}
    \caption{The long time monitoring of cleanliness inside the hall, as well as some key clean control measures are shown in the figure. Some of the larger peaks from March to July in 2023 mainly resulted from the dispersion of acrylic powder into the air during the polishing of the acrylic panel joints.}
    \label{fig:cleanliness}
\end{figure*}

We conducted cleanliness level measurements with a particle size analyzer (Lighthouse Handheld 3016) at multiple points inside the hall and found that the differences were within 10\%. This is primarily due to the continuous operation of four circulating air purification fans with a capacity of 200,000 m$^3$/h, throughout the entire installation period. Their sustained operation promotes effective mixing and circulation of air, leading to a relatively uniform distribution of particulate matter in the hall. Therefore, we selected a fixed point within the hall for long-term stability testing. The long time monitoring of cleanliness inside the hall is shown in Fig.~\ref{fig:cleanliness}. Some key clean control measures have been marked in the figure. According to the particle size distribution in the class 10,000 environment (ISO class 7), all particles can be considered as spherical to obtain the total particle volume per unit volume of air at the class 10,000 level. We continuously monitor the particle size distribution in the air inside the hall using the particle size analyzer  and calculate the real-time total particle volume per unit volume based on spherical particles. The values on the Y-axis in the graph represent the real-time total particle volume divided by the total particle volume at the class 10,000 level, indicating the cleanliness level relative to the class 10,000 level. With all of the efforts described above, the cleanliness inside the experimental hall has reached the equivalent of a class 10,000-100,000 level since May 2022, with an average cleanliness level of class 74,000. The above cleanliness control measures have proven highly effective and can be applied to contamination prevention in other experiments of the similar condition.

The 35.4-meter-diameter acrylic sphere is assembled on-site with 265 acrylic panels, with each layer being meticulously polished to improve its smoothness. Starting in 2023, the installation of acrylic increased the number of construction workers, and the installation schedule was changed to two shifts per day. Some of the larger peaks from March to July in the graph mainly resulted from the dispersion of acrylic powder into the air during the polishing of the acrylic panel joints. During non-polishing periods, the cleanliness level in the hall mostly remains at class 10,000. Since the acrylic material we used to build the acrylic sphere was very clean -- less than 10$^{-12}$~g/g measured U/Th concentrations, the acrylic powder generated during the polishing process has minimal impact on the background compared to rock dust.

\subsection{Detector cleaning and protection}
\label{4.2}
The SS truss installation of the JUNO detector was finished in June 2022: the truss then underwent three extensive cleanings in 2022, including water flushing and wiping processes. Moreover, before installing each of the PMTs layers, we conducted another thorough cleaning of the corresponding area in the SS truss. 

The installation of the acrylic sphere began in July 2022 and was completed by the summer of 2023 at the equatorial layer, with the lower hemisphere installation completed in October 2024. The outer surface of the acrylic sphere was covered with a protective film throughout the installation process. Unlike the film attached to the inner surface, the film on the outer surface is made of polyethylene with a minimal amount of regular adhesive on one side. This film requires manual removal, and laboratory tests have confirmed that the residual background impact of the adhesive is negligible~\cite{Li:2023rae}. From September to October 2024, we completely removed the protective film, cleaned the acrylic outer surface using a clean cloth, and painted the seam with epoxy (radiopurity screened by ICP-MS: $<$~5~ppt $^{238}$U and $<$~7~ppt $^{232}$Th). The inner surface of the acrylic sphere was also protected by another type of film during the construction process, which was ultimately removed by final 3D high-pressure spraying inside the sphere.

The mass installation of PMTs started in early 2023 and reached the equatorial layer by April 2024, with the completion of the lower hemisphere installation in December 2024. In February 2024, we cleaned the surface of the installed PMTs area with an electrostatic duster.

\section{U/Th settling measurement}
\label{sec5}

\begin{figure}[!h]
    \centering
    \includegraphics[width=0.35\textwidth]{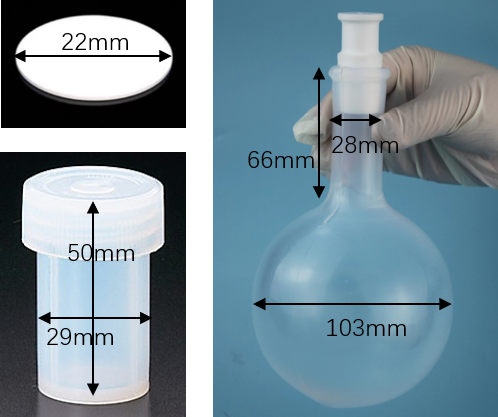}
    \caption{Sampling plate and vessels. The total sampling areas are 4~cm$^2$, 60~cm$^2$, and 390~cm$^2$ for the PTFE plate and acrylic plate, PFA bottle, and PFA flask, respectively. }
    \label{fig:vessel}
\end{figure}

We designed three types of dust-collecting containers: flat plates, a cylinder, and a sphere, as shown in Fig.~\ref{fig:vessel}. The flat plate was intended to examine differences in U/Th deposition rates based on orientation. The cylinder and sphere were designed to simulate geometries commonly found in our detector systems—cylinders resemble tanks often in contact with LS, while the sphere represents the structural form of the JUNO detector. By using these different shapes, we aim to better evaluate U/Th deposition rates across various types of detectors. 

Given that containers made of PFA material typically exhibit an extremely low background, they are well-suited for ultra-low concentration U/Th measurements when combined with ICP-MS. To match the sample processing containers, sampling plates were custom-made only in PTFE and acrylic materials. These were used to compare the U/Th deposition rates between acrylic and PTFE, which, to some extent, reflects the differences in deposition rates on the surfaces of PFA and acrylic glass materials. 

The plates were placed in different orientations (upward, downward, and sideways) to compare the U/Th settling rates in different directions. The idea of this measurement is similar to that carried out at Pacific Northwest National Laboratory and the SNOLAB underground facility~\cite{diVacri:2020aqc}. The PFA bottle simulates the settling rate on the surface of a cylindrical detector, while the PFA flask simulates the settling rate on the surface of a spherical detector, which is the form of the JUNO acrylic sphere. 

\subsection{Plates and containers cleaning}
All the sampling plate and containers underwent the following rigorous cleaning process to minimize the external contamination: 
\begin{itemize}
    \item Degreasing: all sampling plates and containers are soaked in a 0.1\% Alconox (Alconox, Inc.) solution and sonicated for 5~min, followed by rinsing with deionized water (produced by the Milli-Q equipment), then sonicated in deionized water for another 5~min. 
    \item Primary acid washing: the sonicated sampling plates and containers are placed in an acid tank containing 20\% nitric acid (GINUIS) and soaked overnight. 
    \item Secondary acid washing: after being removed from the acid tank and rinsed with deionized water, the sampling plates and containers are boiled in 20\% GINUIS nitric acid for 30~min and then rinsed with deionized water.
\end{itemize}

The sampling plates and containers will then undergo a blank test, as detailed in Section~\ref{5.3}. They are approved for use only if the $^{238}$U/$^{232}$Th concentrations measured in the acid solution are consistent with the average blank value.

\subsection{Sampling}
\begin{figure*}[!h]
    \centering
    \includegraphics[width=0.8\textwidth]{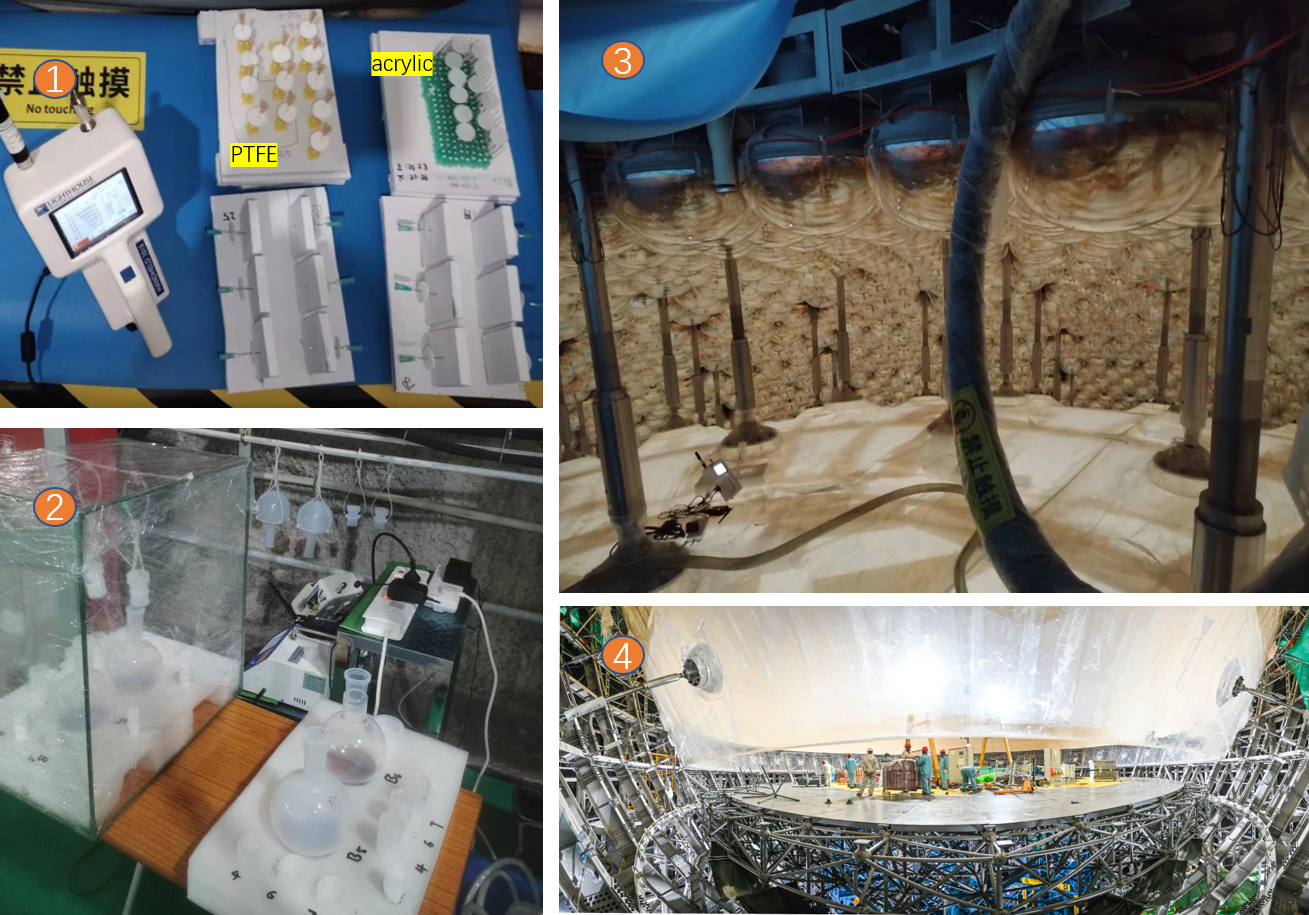}
    \caption{Some sampling pictures in JUNO experimental hall. Fig.~1 and Fig.~2 depict the sampling photographs taken from the installation room and the hall floor, respectively. Fig.~3 shows the sampling image captured between the layers of acrylic sphere and the PMTs. Figure 4 illustrates the installation platform for the acrylic sphere, where we conducted our sampling.}
    \label{fig:sampling}
\end{figure*}

Each plate has a single-side surface area of approximately 4~cm$^2$, so six such sides give a total surface area of 23~cm$^2$. To sample these surfaces in different orientations with the same sampling area, we use the following approach:
\begin{itemize}
    \item For side sampling, a thin needle is used to pierce the sheet and secure it onto a dust-free plastic pad, allowing both sides to collect dust. Thus, only three sheets are needed for one sample in this orientation.
    \item For upward- and downward-facing sampling, to prevent the non-sampling side from collecting dust, two sheets are stacked together and fixed with a thin needle onto a dust-free plastic pad. This means six sheets are required for upward-facing and six for downward-facing sampling, totaling twelve sheets.
\end{itemize}
Therefore, a full set of samples requires 15 sheets in total. For each sampling location, we collect 2–3 parallel samples. Figure~\ref{fig:sampling}.(1) shows two parallel samples for each of the two types of sampling sheets.

Both the PFA bottle and PFA flask are sampled with the caps kept open, with both the bottle mouth and the cap facing upwards or downwards (Fig.~\ref{fig:sampling}.(2)). To investigate whether there are differences in the settling of U/Th in open and closed spaces, we created a simple small cover and placed the PFA sampling bottles inside the cover for sampling, comparing it with the sampling bottle outside the cover (Fig.~\ref{fig:sampling}.(2)). 

There are a total of four sampling points in the JUNO underground experimental halls, including the Top Hall, the PMTs and acrylic interlayer (Fig.~\ref{fig:sampling}.(3)), the acrylic installation platform (Fig.~\ref{fig:sampling}.(4)), and the installation space outside the main hall. Each sampling point is equipped with three types of sampling containers and a particle size analyzer (Lighthouse Handheld 3016), which is used for continuously monitoring the particle size distribution.

\subsection{Sample treatment process}
\label{5.2}
The acrylic sphere was assembled in the JUNO experimental hall by bonding it layer by layer. After each layer of acrylic is assembled, the seams need to be polished to improve the smoothness. During the polishing process, a large amount of acrylic powder is generated, which floats in the air and is collected in our containers. Therefore, the sample treatment process not only requires the use of acid to dissolve the inorganic salts, but also requires the removal of the acrylic powder inside through filtration or ashing methods. 

Both the PTFE and acrylic plates can fit into a 23~mL PFA bottle, so the sampling and sample processing are both done in the PFA bottle. The sample treatment process for all samples is as follows: 

\begin{itemize}
    \item Add 4~mL (20~mL) of 5\% nitric acid (Fisher Scientific) and 0.5~g of hydrofluoric (HF) acid (Fisher Scientific) to a PFA bottle (PFA flask), and boil it for 25~min. The purpose of adding HF acid here is to dissolve any silicates that may be present in the sample.  
    \item For the plates, after acid boiling, the plate is removed and rinsed with 1~mL of boiled 5\% nitric acid to ensure that any residual sample solution on the plate is also washed off. 
    \item Remove organic powder: in order to remove the acrylic powder from the acid solution, the acid solution needs to be either filtered or ashed. For the filtration method, all acid solutions are collected using a syringe and filtered through a membrane with a pore size of 0.2~$\upmu$m. The filtered liquid is subsequently evaporated to dryness to eliminate HF acid, which could otherwise corrode the ICP-MS. Thereafter, 2~g of 20\% nitric acid is added, and the mixture is boiled to dissolve the residue. For the ashing method, all acid solutions are transferred to a platinum crucible and placed in a muffle furnace, where they are completely ashed at 600$^{\circ}$C. Then, 5~g of 20\% nitric acid is added and boiled to dissolve the inorganic residue retained in the crucible. We got similar results from these two methods of treatment, so we chose filtration for future samples.
    \item Measurement: after the acid is made up to volume (2~mL), the sample is sent for ICP-MS measurement.
\end{itemize}

\subsection{Blanks and recovery efficiency}
\label{5.3}
The blank experiment involves repeating all the pre-treatment processes for the samples, the only difference being that no sample is added. The blank results can be used to assess the level of external contamination introduced during the entire pre-treatment process, including contributions from the vessel surfaces and ambient air. Subsequently, a fixed volume of 2~mL of the acid solution is transferred to the ICP-MS for analysis. By multiplying the measured concentration obtained from ICP-MS by the 2~mL solution volume, the exact mass of contamination attributable to that vessel and procedural step can be determined. The blank results of the pre-treatment process are summarized in Table~\ref{tab:blank}. The PTFE is not as clean as PFA, so the blanks are above 1~pg with relatively larger standard deviation (SD). The method detection limit (MDL) at 99\% confidence level (C.L.) is calculated as $\overline{X}+t\times SD$, where $\overline{X}$ is the average blank on U/Th, $t$-value relies on the number of replicates~\cite{MDL}. From the results, it can be seen that the U/Th blanks of PFA bottle, PFA flask, and platinum crucible after acid boiling and filtration operations are 0.2-0.6~pg, while the blanks of PTFE plate and acrylic plate acid boiling and ashing are 1-2~pg. 

\begin{table*}[htbp]
\centering
 \caption{ \label{tab:blank} Blank results for different treatments.}
\resizebox{0.8\textwidth}{!}{%
 	\begin{tabular}{c|c|c|cc|c|cc|c}
        \hline
    \multirow{2}{*}{Vessel} &  \multirow{2}{*}{Pre-treatment} &  \multirow{2}{*}{Parallels} & \multicolumn{3}{c|}{$^{238}$U [pg]} & \multicolumn{3}{c}{$^{232}$Th [pg]} \\ \cline{4-9}
    &&& $\overline{X}$ & $SD$ & MDL & $\overline{X}$ & $SD$ & MDL \\ \hline
    PFA bottle & acid boiling+filter & 24 & 0.2 & 0.1 & 0.4 & 0.2 & 0.1 & 0.4 \\ \hline
    PFA flask & acid boiling+filter & 16 & 0.3 & 0.1 & 0.6 & 0.5 & 0.2 & 0.9 \\ \hline
    Platinum crucible & acid boiling+ashing & 9 & 0.4 & 0.3 & 1.3 & 0.6 & 0.3 & 1.6 \\ \hline
    PTFE plate (6 pieces) & acid boiling+filter & 10 & 0.8 & 0.4 & 2.0 & 1.1 & 0.7 & 3.0 \\ \hline
    Acrylic plate (6 pieces) & acid soaking+filter & 8 & 1.5 & 0.6 & 3.3 & 2.2 & 1.7 & 7.3 \\ 
    \hline 
    \end{tabular}
    }
\end{table*}

All sampling plates and containers must undergo this blank test. Only those that yield normal results are approved for use in sampling. 

\begin{figure}[!h]
    \centering
    \includegraphics[width=0.48\textwidth]{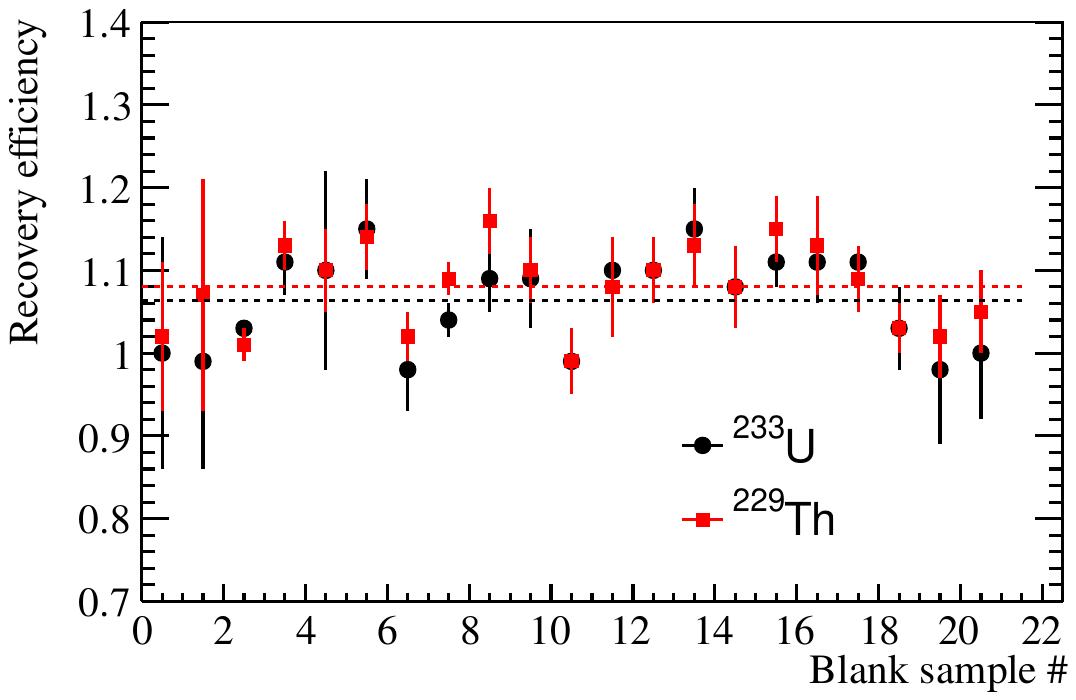}
    \caption{The U/Th recovery efficiency for the entire pre-treatment. The average efficiencies of $^{233}$U and $^{229}$Th are shown in the black and red dash line.}
    \label{fig:eff}
\end{figure}

The entire pre-treatment process's U/Th recovery efficiency was determined by adding known amounts of naturally non-existent $^{229}$Th and $^{233}$U standards (preserved in 5\% nitric acid) to the samples, subjecting them to the same pre-treatment process, and finally sending them for ICP-MS testing. The measurement results of 10 samples showed U/Th average recovery efficiency around 106\%/108\%, with a standard deviation of 5\%, as shown in Fig.~\ref{fig:eff}. The possible reason for a recovery rate exceeding 100\% is that the sampling and sample processing containers are reused in the laboratory. During the sample processing, U/Th elements can permeate into the container walls and also leach out from them. After the cleaning procedure described in Section~\ref{5.2}, the containers reach an equilibrium between the permeation and leaching processes during sample handling. This equilibrium is corrected for the ratio of $^{232}$Th and $^{238}$U by using the recovery rates of $^{229}$Th and $^{233}$U.

\subsection{Results}
Outside the JUNO underground experimental hall, there is an installation room with a cleanliness level of class 40,000-50,000. The measurement data collected from all sampling containers placed at the same location (installation room) are presented in Table~\ref{tab:result1}, aiming to eliminate the interference of acrylic powder in the experimental hall. For samples, the U/Th concentration in the 2~mL solution is similarly measured via ICP-MS. Multiplying this concentration by the 2~mL volume yields the total mass of U/Th present in the solution. Subtracting the corresponding blank mass (from Table~\ref{tab:blank}) gives the net deposited mass of U/Th originating from the sample itself. This net mass is then divided by the sample's surface area and exposure duration to calculate the deposition rate (pg/d/m$^2$).

From the results in the table, the following conclusions can be drawn: 

\begin{itemize}
    \item The settling rate of U/Th in open spaces (No.2 and No.5) is about two orders magnitude higher than in closed spaces (No.1 and No.4);
    \item The settling rate of U/Th in spherical flasks (No.5) is about six times lower than in cylindrical bottles (No.2);
    \item For cylindrical bottles (No.2 and No.3) and spherical flasks (No.5 and No.6), the settling rate of U/Th with the mouth of the bottle facing upwards is 10-40 times higher than when the mouth is facing downwards.
    \item The chimney height of the JUNO detector is 9~m, and the diameter of the sphere is 35.4~m. Therefore, the ratio of the chimney height to the sphere diameter is 25\%. The ratio of the upper cylindrical section to the lower spherical diameter of the PFA flask is 64\%. Concerned about whether the upper straight section might affect the settling inside the sphere, we conducted the following comparative experiment: one original PFA flask and one modified PFA flask with the upper straight section cut down to only 1~cm in height, resulting in a ratio of the upper cylindrical section to the lower spherical diameter of 10\%. We compared whether there was any difference in the settling rate at the same position between the original PFA flask and the shortened PFA flask. The results (No.~5 and No.~7) showed that the U/Th sedimentation outcomes of both flasks were consistent within the uncertainty. 
    \item The deposition amount of U/Th on the upper surface of the PTFE plate (No.8) is similar to that on the acrylic plate (No.11), but the results show that the deposition amount on the lower surface (No.9) or side (No.10) of the PTFE plate is 3-20 times higher than that of the acrylic plate (No.12 and No.13). The deposition on the upper surface is a combination of dust settling and adsorption processes. From the results, the deposition on the upper surface is primarily due to gravitational settling of dust, while the deposition on the lower surface or side is mainly attributed to adsorption.
\end{itemize}

\begin{table}[htbp]
\centering
 \caption{ \label{tab:result1} All sampling containers are placed in the installation room for 12-14 days in May-Sep 2024. The average values and standard deviations for $^{238}$U and $^{232}$Th settling computed with sample size of three parallels samples are shown in the last column. The PFA flask with label "1~cm" represents the modified PFA flask with the upper straight section cut down to only 1~cm in height. }
 \resizebox{0.5\textwidth}{!}{%
 	\begin{tabular}{c|c|cc|cc}
        \hline
    \multirow{2}{*}{No.} & \multirow{2}{*}{Vessel} & \multirow{2}{*}{Space} & \multirow{2}{*}{Direction}  & \multicolumn{2}{c}{Settling [pg/d/m$^2$]} \\\cline{5-6}
    &&&& $^{238}$U & $^{232}$Th \\ \hline
    1& \multirow{3}{*}{PFA bottle} & closed & upwards & $<$3.0 & $<$3.6 \\ 
    2&& open & upwards & 280$\pm$15 & 438$\pm$26 \\ 
    3&& open & downwards & 10$\pm$2 & 12$\pm$2 \\ \hline
    4&\multirow{3}{*}{PFA flask} & closed & upwards & $<$0.7 & $<$1.0 \\ 
    5&& open & upwards & 51$\pm$14 & 73$\pm$23 \\ 
    6&& open & downwards & 4$\pm$1 & 2.0$\pm$0.6 \\ \hline
    7 & PFA flask (1~cm) & open & upwards & 43$\pm$17 & 54$\pm$26 \\  \hline
    8&\multirow{3}{*}{PTFE plate} & open & upwards & 1624$\pm$232 & 1869$\pm$67 \\
    9&& open & downwards & 394$\pm$169 & 504$\pm$345 \\
    10&& open & sideways & 406$\pm$98 & 371$\pm$25 \\  \hline
    11&\multirow{3}{*}{Acrylic plate} & open & upwards & 1166$\pm$442 & 1809$\pm$887 \\
    12&& open & downwards & 24$\pm$6 & 93$\pm$83 \\
    13&& open & sideways & 157$\pm$60 & 98$\pm$9 \\ 
    \hline 
\end{tabular}
    }
\end{table}

At about the same time, we placed the sampling bottles and particle counting machine at different locations within the JUNO experimental hall, and left them collect the dust with the same exposure period settled for the outside installation room. The measured settling rates of U/Th and cleanliness levels are summarized in Table~\ref{tab:result2}.

\begin{table}[htbp]
\centering
 \caption{ \label{tab:result2} Sampling was conducted using the same type of sampling bottle and facing upwards at different locations for 9-14 days in April-Sep 2024. The average values and standard deviations for $^{238}$U and $^{232}$Th settling computed with sample size of three parallels samples are shown in the last column.}
 \resizebox{0.5\textwidth}{!}{%
 	\begin{tabular}{c|c|cc|cc}
        \hline
    \multirow{2}{*}{No.} & \multirow{2}{*}{Vessel} & \multirow{2}{*}{Location} &  Cleanliness  & \multicolumn{2}{c}{Settling [pg/d/m$^2$]} \\ \cline{5-6}
    &&& [$\times 10^4$]& $^{238}$U & $^{232}$Th \\ \hline
    1& & Top Hall  & 4.8 & 691$\pm$60 & 926$\pm$26 \\ 
    2& PFA & Interlayer & 13 & 397$\pm$11 & 445$\pm$71 \\ 
    3& bottle & Platform & 216 & 657$\pm$72 & 804$\pm$49 \\ 
    4&& Installation hall & 3.8 & 280$\pm$15 & 438$\pm$26 \\ \hline
    5& PFA & Top Hall & 4.8 & 67$\pm$43 & 77$\pm$48 \\ 
    6& flask & Interlayer & 13 & 34$\pm$11 & 34$\pm$19 \\ 
    7& & Installation hall & 3.8 & 51$\pm$14 & 73$\pm$23 \\ \hline
\end{tabular}
}
\end{table}

From the results of the PFA bottle, the U/Th settling rates in the Top Hall (No.~1) and the acrylic installation platform (No.~3) are similar; however, the air particle counting differs by an order of magnitude. This discrepancy is due to the fact that the acrylic installation process involves polishing the surface of the bonding area, resulting in a significant amount of acrylic powder that interferes with the measurement of airborne dust particle sizes. The U/Th settling rates in the acrylic and PMTs interlayers are close to those measured outside the hall, likely because there is less human activity in this area compared to the Top Hall and installation platform. For the flask container, the geometric effects lead to a settling rate that is nearly an order of magnitude lower than that of the PFA bottle.

\section{Application to the JUNO detector systems}
\label{sec6}
Background contributions to the JUNO detector from operational and environmental contamination during installation may arise through the following mechanisms:

\begin{itemize}
\item The inner surface of the acrylic vessel was protected by film throughout most of the installation process. However, in some areas, the film may have detached, exposing the surface to air and resulting in deposition of natural radionuclides. Furthermore, during the final flushing process, residual adhesive film and particulate impurities that were not completely removed could contaminate the LS.
\item The purification and filling systems associated with the LS involve extensive use of pipelines, tanks, connectors, and instrumentation. These components underwent dedicated cleaning and inspection procedures to ensure surface cleanliness. Similar to the acrylic cleaning process, residual impurities that were not fully removed could potentially contaminate the LS. Furthermore, after cleaning, exposure to air during installation resulted in the deposition of additional impurities on the surfaces.  During LS production and transfer, these contaminants may be flushed into the LS.
\item Detector components located outside the acrylic vessel that do not directly contact the LS—such as structural supports and external modules—may still contribute to the background. Although they do not directly contaminate the LS, gamma rays emitted from natural radionuclides within these materials can penetrate the water shielding and the acrylic vessel, depositing energy in the LS.
\end{itemize}

The following sections will quantitatively assess the impact of each of these three pathways based on available measurement data. 

\subsection{Acrylic inner surface}
After the acrylic sphere was constructed, a 3D spray head was placed into the sphere from the top. The first step is to spray UPW mist to reduce the dust particle content inside the sphere, with a cleanliness improvement to class~1,000 level. The second step involves high-pressure UPW flushing of the inner surface of the acrylic sphere to remove the surface protective film and perform the final cleaning. It is challenging to directly measure the level of contaminants remaining on the acrylic surface after flushing. Instead, the effectiveness of the cleaning process was evaluated by monitoring the particulate content in water samples discharged from the bottom outlet. Flushing was continued until the particle concentration in the water consistently met the Level 50 standard (MIL-STD-1246C)\footnote{Cleanliness Level 50 standard of MIL-STD-1246C specifies a maximum particle count per liter of: 530 for particles $\ge$5~$\upmu$m, 230 for particles $\ge$15~$\upmu$m, 34 for particles $\ge$25~$\upmu$m, and 10 for particles $\ge$50~$\upmu$m.} and stabilized. As a result, the amount of residual contaminants that were not removed from the acrylic surface remains undetermined. The actual cleanliness of the acrylic surface can only be verified through experimental data. After completing these two steps, the cleanliness level inside the acrylic sphere was measured at various heights and found to range between Class 100 and Class 1,000.

Upon completion of the cleaning process, the upper chimney was sealed with an acrylic blind plate, and the lower chimney was fitted with a SS cover to prevent extensive air exchange between the interior and exterior of the sphere. Subsequently, the remaining 8-meter section of the SS chimney and the calibration room were installed. Each section was sealed with nitrogen gas after installation. Once the calibration room was fully installed, the blind plate on the upper chimney was removed, and nitrogen gas was circulated through the chimney to maintain a positive pressure within both the upper chimney and the calibration room.

The total area of the acrylic inner surface amounts to 4,000~m$^2$. The acrylic filling process involves initially filling the inside and outside of the sphere with UPW, then displacing the UPW with LS inside the sphere. The entire time for cleaning took 15~days, and after additional 15~days for chimney and pipes installation, we started the UPW filling of the vessel on Dec 18, 2024. Based on the experimental results from the previous section, the settling rate of U/Th in spherical flasks during exposure in a closed space and controlled environment of class~1,000 is estimated to be lower than 0.02~pg/m$^2$/d, assuming a linear change in settling rate with cleanliness and similar U/Th settling on PFA flask and acrylic sphere. Therefore, the estimated mass of U/Th settling on the entire acrylic inner surface during the one-month pre-filling exposure period (assuming a cleanliness level of Class 500-5,000) is approximately 1.3 to 13 ng. During the filling process, the acrylic sphere and water surface were also exposed to air, but the exposed acrylic surface would change with the filling process. The entire filling process took two months, and the average acrylic surface exposure is one month due to the filling process. Therefore, the estimated U/Th settling during filling period is similar to that of before filling. Assuming all U/Th settling on the surfaces of the acrylic sphere and water enters the 20~kt LS, it would contribute to 0.13-1.3$\times$10$^{-18}$~g/g of U/Th in LS, which is well below our requirement. 

One day before the completion of the acrylic sphere cleaning, approximately 6~m$^2$ (0.15\% of the whole acrylic inner surface) of protection film remained attached to the inner surface of the upper hemisphere, as checked by a person from the outer surface of acrylic sphere on Dec.~1$^{st}$. This is mainly due to the epoxy adhesive at the bonding line potentially sticking to the water-soluble paper while it was not completely dry, resulting in incomplete washing away by the water. Most of the residual paper is located on the upper hemisphere of the sphere, primarily for two reasons: Firstly, during the assembly process of the lower hemisphere, the paper film at the seams is applied in reverse, meaning the adhesive side faces outward to prevent it from bonding with the epoxy. Secondly, all the seams on the lower hemisphere are removed during the final inspection process inside the sphere after its construction. It is worth noting that the amount of residue mentioned here is a rough estimate made by workers through visual inspection, and the results may be slightly exaggerated, as working on the outer surface of the sphere is relatively challenging. The radiopurity of the paper film was screened using a high-purity germanium detector, revealing activity levels of $<$0.5~Bq/kg for $^{238}$U ($^{214}$Pb and $^{214}$Bi gammas) and 0.6$\pm$0.2~Bq/kg for $^{232}$Th ($^{212}$Pb, $^{228}$Ac, and $^{208}$Tl gammas) by assuming the chain is at an equilibrium state, based on a 300~g sample. With a density of 47~g/m$^2$, a residual of 0.15\% (equivalent to 6~m$^2$) would result in approximately 280~g of film remaining on the inner surface of the acrylic. 

To assess the contamination level caused by the water-soluble paper residual upon contact with the LS, we conducted an aging experiment by soaking 25~cm$^2$ paper film in 60~g LS at 80$^{\circ}$C for 24 hours. For the aging test, a 10$^{\circ}$C increase in temperature yields an acceleration factor of approximately 2.5~\cite{IEC61709:2017}, with the activation energy parameter set to about 80~kJ/mol. Consequently, soaking the paper film in LS at 80$^{\circ}$C for 24 hours is equivalent to aging at room temperature (21$^{\circ}$C) for approximately 244 days. Subsequently, we measured the U/Th content in the LS before and after soaking by acid extraction and ICP-MS~\cite{Li:2024gqe}. The results were shown in Table~\ref{tab:soaking}, and we found that the U/Th leaching effect from the water-soluble paper was at a level of 200-400~pg/m$^2$/y. Considering the 0.15\% of paper residual and 20~kt LS mass, the final U/Th extraction from film to LS is 0.6-1.3$\times$10$^{-19}$~g/g per year, which is significantly lower than the design specifications and expected runtime of JUNO. As for the single-event background caused by residual paper left on the acrylic inner surface, the measured activity of the 280~g residual film is approximately 0.2~Bq, with a U/Th concentration level of 0.5–0.6~Bq/kg. This background can be effectively removed through the fiducial volume cut in the analysis.

\begin{table}[htbp]
\centering
 \caption{ \label{tab:soaking} The ICP-MS screening results of U/Th content in the LS before and after film soaking are shown in the table. The U/Th extraction rate in the last column is subtracted blank of U/Th without film soaking.}
 \resizebox{0.5\textwidth}{!}{%
	\begin{tabular}{c|c|c|c|c|c}
        \hline
        \multirow{3}{*}{} &  \multicolumn{2}{c|}{Without film [pq]} & \multicolumn{2}{c|}{With film [pg]} & Extraction rate   \\ \cline{2-6}
        & No.~1 & No.~2 & No.~1 & No.~2 & [pg/m$^2$/y] \\ \hline
        $^{238}$U & 0.6$\pm$0.2 & 0.9$\pm$0.2 &  5.2$\pm$0.8 & 5.3$\pm$0.8 & 193$\pm$21  \\    
        $^{232}$Th & 0.3$\pm$0.2 & 0.7$\pm$0.1 & 12$\pm$1 & 11.0$\pm$0.5 & 420$\pm$41 \\ \hline
    \end{tabular}
}
\end{table}

\subsection{LS purification and filling systems}
The JUNO LS is subjected to a sequential four-stage purification process, which includes alumina adsorption, distillation, water extraction, and gas stripping. After this process, it is filled into the acrylic sphere by a filling system. The primary objectives of the alumina and gas stripping systems are to enhance the transparency of the alkylbenzene feedstock and to remove gaseous components from the LS, respectively. Consequently, their capabilities for the purification of U/Th are somewhat limited. The efficiency of the distillation system~\cite{Landini:2024ohd} and water extraction system~\cite{Ye:2021jkp} in removing U/Th impurities is expected to be 99\% and 80\% respectively. The total inner surface area of the entire system, as well as the surface area weighted by purification efficiency, are summarized in Table~\ref{tab:LS}. With U/Th contamination in the 20~kt LS reaching levels of 10$^{-18}$~g/g, corresponding to a U/Th mass of 20~ng. Therefore, for the 1105~m$^2$ inner surface area of the LS-related system, the required U/Th residual should be smaller than 18~pg/m$^2$. 

Most pipes, containers, and components in the LS-related systems that are acid-resistant underwent degreasing, acid pickling, passivation, and UPW rinsing. The final validation of cleanliness included the following criteria: 
\begin{itemize}
\item The particle concentration in the rinse water must meet Level 50 standard (MIL-STD-1246C). 
\item The change in resistivity before and after UPW rinsing should remain within 4–6 M$\Omega\cdot$cm. 
\item There is no significant change in the absorption spectrum of the rinse water before and after rinsing.
\item U/Th concentrations in the rinse water must be below 10$^{-13}$~g/g. 
\end{itemize}
Only when all these criteria are met is the component considered qualified. Similar to the challenges in acrylic surface cleaning, the exact amount of residual contamination on cleaned surfaces remains unknown. Therefore, we implemented multiple verification methods to ensure the effectiveness of the cleaning process. 

After cleaning, all these components were sealed with nitrogen. During installation, however, these systems may be exposed to ambient air for limited periods. The acceptable duration of such exposure can be estimated based on the measurement results presented in the previous section. The cleanliness of the room housing the underground LS purification and filling systems has also been monitored long-term, maintaining a class 1,000-10,000 standard. These systems are mostly cylindrical, so we plan to use the measurement results from PFA bottles for background evaluation. Based on the results from PFA bottle No.2 (facing upwards) in Table~\ref{tab:result1}, the U/Th settling rate in a Class 10,000 environment is estimated to be approximately 100 pg/d/m², scaled according to the cleanliness level. This estimate accounts for an exposure time of 4.3 hours under Class 10,000 conditions. During construction, all interfaces of pipelines and containers are covered with blind plates before docking, and the blind plates are removed during docking. The vast majority of interfaces strictly require an air exposure time during docking of less than 10~min, with internal nitrogen gas protection after docking is completed. Therefore, even if the environmental cleanliness level fluctuates several times higher than the class 10,000 level, the impact of air exposure during the installation of the LS-related system is controllable.

\begin{table}[htbp]
\centering
 \caption{ The weighted surface area considering the 99\% and 80\% efficiency from distillation and water extraction.\label{tab:LS} }
 \resizebox{0.5\textwidth}{!}{%
 	\begin{tabular}{c|c|c|c}
        \hline
        & Surface area [m$^2$] & Removal eff. & Weighted area [m$^2$] \\ \hline
        Before distillation & 2500 & 99\% & 25 \\
        Before extraction & 900 & 80\% & 180\\
        After extraction & 900 & 0 & 900\\ \hline
        \multicolumn{3}{c|}{Total} & 1105\\ \hline
\end{tabular}
}
\end{table}

\subsection{Structure outside the acrylic}
The dust deposited on the outer surface of the acrylic sphere, PMTs, and SS truss during the construction process, can produce gamma rays due to its natural radioactivity. The gamma rays can penetrate through the water layer and the acrylic sphere and enter the LS, which contributes to the background. Based on the JUNO software framework~\cite{Lin:2022htc}, we simulated the natural radioactivity on the outer surface of the acrylic sphere, PMTs, and the SS truss~\cite{JUNO:2021kxb}. The probability for the deposited energy of being greater than 0.7~MeV and be positioned within the fiducial volume (R $<$ 17.2~m) of the LS for each natural radioactive nuclide is listed in Table~\ref{tab:MCeff}. For $^{238}$U and $^{232}$Th,
the full decay chain was simulated. Taking the acrylic outer surface as an example, 1~Bq of $^{238}$U will lead to 0.001~Hz event rate in the LS with deposited energy larger than 0.7~MeV at a position with radius smaller than 17.2~m, while the rate is one order of magnitude larger for $^{232}$Th, which has more energetic gamma rays in the chain ($^{208}$Tl). At the same time, the radon produced by $^{238}$U in the dust will enter the shielding water, causing an increase of radon concentration. 

\begin{table}[!h]
\begin{center}
\footnotesize
\renewcommand\arraystretch{1.3}
 \caption{ \label{tab:MCeff} The full decay chain of $^{238}$U and $^{232}$Th on the outer surface of the acrylic sphere, PMTs, and the SS truss were simulated based on the JUNO software framework. The probability of deposited energy being greater than 0.7~MeV and position within the fiducial volume (R $<$ 17.2~m) of LS from each radioactivity is listed in the table.}
 	\begin{tabular}{c|c|c|c}
        \hline
    \multicolumn{2}{c|}{} & Acrylic outer surface & PMTs$\&$Truss  \\ \hline
    \multicolumn{2}{c|}{Radius [m]} & 17.8 & 19.2-20.0 \\ \hline
    \multirow{2}{*}{Probability} & $^{238}$U & 1.1$\times$10$^{-3}$ & 3.4$\times$10$^{-6}$ \\ \cline{2-4}
    & $^{232}$Th & 1.3$\times$10$^{-2}$ & 1.4$\times$10$^{-5}$ \\     \hline 
\end{tabular}
\end{center}
\end{table}

As shown in Table~\ref{tab:result1}, for flat plates made of PTFE and acrylic, the results for the upward-facing orientation are quite similar ($>$~1000~pg/d/m$^2$). Although there are differences in the results for the downward-facing and side-facing orientations, their contributions are relatively minor ($\leq$~500~pg/d/m$^2$) compared to the upward-facing results. A similar trend is observed for PFA bottles and flasks. This indicates that gravitational settling of dust during environmental exposure plays a dominant role. Therefore, even though the previous experiments did not simulate all detector material types, we can still refer to the results from PFA bottles and flasks when evaluating deposition rates. Since the aforementioned detector structures are uniformly spherical, we intend to use the measurement results from PFA flasks for background evaluation. 

The installation situation and progress for acrylic sphere, PMTs and SS truss are summarized in Table~\ref{tab:juno}, as well as the expected U/Th settling rate. Details on the construction can be found in Section~\ref{4.2}. The exposure here only include the time after the final cleaning and before filling. The U/Th settling used in the calculation is the results of PFA flask from Table~\ref{tab:result2} during the exposure with the sphere facing upwards (No.~5 for acrylic and truss, No.~6 for PMTs layer). Based on the U/Th efficiency-to-singles-rate ratio within the FV for energies above 0.7~MeV from simulation (Table~\ref{tab:MCeff}), the single-event rate originating from deposited U/Th on the detector surface inside the LS has been calculated and is presented in the penultimate row of the table. Taking into account the 9-kt UPW volume between the acrylic vessel and the truss, the rate of radon release from deposited U/Th of the detector surface to the water has also been calculated and is shown in the final row. Taking the acrylic column as an example, the total activity of $^{238}$U is 0.37~Bq. Assuming secular equilibrium, the activity of $^{222}$Rn is also 0.37~Bq. Dividing this by the volume of the inner water layer (9$\times$10$^3$~m$^3$) gives a radon concentration of 0.04~mBq/m$^3$. Overall, the U/Th contamination introduced during the installation of the detector due to environmental exposure contributes approximately 2~mHz to the single-event rate in the LS fiducial volume (R $<$ 17.2~m), accounting for about 0.03\% of the total detector contribution (7~Hz~\cite{JUNO:2021kxb}). The impact of radon released from the decay chain products of $^{238}$U deposited on the detector surface into the water is at a level of 2~mBq/m$^3$ by assuming equilibrium rate as $^{238}$U, which is approximately 20\% of the water radon target (10~mBq/m$^3$). Additionally, some of the deposited dust can be further purified through the water circulation process.

\begin{table}[htbp]
\centering
 \caption{ The anticipated accumulation of U/Th in the JUNO main detector components is calculated based on exposure time and the results of U/Th fallout measurements. The potential singles rate in the LS fiducial volume (FV, R $<$ 17.2~m) and the radon contamination in water are presented in the last two rows. \label{tab:juno} }
 \resizebox{0.5\textwidth}{!}{%
 	\begin{tabular}{c|c|cc}
        \hline
    & Acrylic & PMTs & SS truss \\ \hline
    Installation start & 2022-07 & 2023-02 & 2022-01 \\
    Install to equator & 2023-08 & 2024-04 & 2022-04 \\
    Installation complete & 2024-10 & 2024-11 & 2022-06 \\
    Final cleaning & 2024-09 & 2024-02 & 2022-12 \\ \hline
    Surface area [m$^2$] & 4e3 & 3e4 & 1.6e4 \\ 
    Exposure [month] & 3 & 9 & 24 \\ 
    Settling [pg/m$^2$/d] & 80 & 35 & 80  \\ 
    Deposited U/Th [$\upmu$g] & 30 & 280 & 920 \\ \hline
    Activity of U/Th chain [Bq] & 0.37/0.12 & 3.4/1.1 & 11.3/3.7 \\ 
    U/Th efficiency from Table~\ref{tab:MCeff} & (1.1/13)$\times$10$^{-3}$ & \multicolumn{2}{c}{(3.4/13)$\times$10$^{-6}$} \\ \hline
    Singles in FV [mHz] & 2 & 0.03 & 0.09 \\
    Radon in water [mBq/m$^3$]& 0.04 & 0.4 & 1.3 \\ \hline
\end{tabular}
}
\end{table}

\section{Summary}
\label{sec7}
The installation of the JUNO detector spanned three years. To ensure the detector's cleanliness, the on-site installation environment was managed under clean-room conditions, with an average cleanliness level of class 74,000. After the installation of the acrylic sphere and before the filling with LS, the first step of the cleaning involved a water mist dust reduction inside the sphere, which improved the air cleanliness to a level of class 100 to 1,000. The second step involved using a 3D high-pressure nozzle to rinse the inner surface of the acrylic sphere, further enhancing the cleanliness of the surface in contact with the LS. It should be noted that both the purification and filling systems in contact with the LS and the inner surface of the acrylic vessel underwent rigorous cleaning and validation procedures to achieve the highest attainable level of surface cleanliness. However, the amount of residual impurities remaining on these surfaces and the extent to which they may contaminate the LS remain unknown and can only be evaluated through actual measurement data.

Surface deposition experiments for U/Th on various containers were conducted in the JUNO experimental hall. Based on experimental results, the total U/Th deposition on the acrylic vessel, PMTs, and SS truss during the detector installation process has been evaluated. Using detector simulations, the efficiency-to-singles-rate ratio for U/Th within the FV at energies above 0.7~MeV was found to be 0.1\% for $^{238}$U and 1.3\% for $^{232}$Th originating from the acrylic surfaces. In contrast, the corresponding efficiency for contributions from the PMTs and SS truss is approximately three orders of magnitude lower due to the presence of about 3~m of water shielding. Consequently, the single-event rate in the LS FV resulting from deposited surface dust accounts for only about 0.03\% of the total rate (7~Hz) arising from all detector materials. The radon contamination level in the inner water region (approximately 9~kt) between the acrylic vessel and the SS truss, originating from U/Th deposits on the surfaces, is approximately 2~mBq/m$^3$. This value meets only 20\% of the maximum permitted limit of 10~mBq/m$^3$.

\section*{Acknowledgements}
This work is supported by the National Key Research and Development Program of China (2024YFE0110503), the Youth Innovation Promotion Association of the Chinese Academy of Sciences, CAS Project for Young Scientists in Basic Research (YSBR-099). 

\bibliographystyle{unsrt}
\bibliography{reference}

@article{JUNO:2021vlw,
    author = "Abusleme, Angel and others",
    collaboration = "JUNO",
    title = "{JUNO physics and detector}",
    eprint = "2104.02565",
    archivePrefix = "arXiv",
    primaryClass = "hep-ex",
    doi = "10.1016/j.ppnp.2021.103927",
    journal = "Prog. Part. Nucl. Phys.",
    volume = "123",
    pages = "103927",
    year = "2022"
}

@article{Lin:2022htc,
    author = "Lin, Tao and others",
    title = "{Simulation software of the JUNO experiment}",
    eprint = "2212.10741",
    archivePrefix = "arXiv",
    primaryClass = "hep-ex",
    doi = "10.1140/epjc/s10052-023-11514-x",
    journal = "Eur. Phys. J. C",
    volume = "83",
    number = "5",
    pages = "382",
    year = "2023",
    note = "[Erratum: Eur.Phys.J.C 83, 660 (2023)]"
}

@misc{MDL,
    author = "United States Environmental Protection Agency 821-R-16-006",
    howpublished = "https://www.epa.gov/sites/default/files/2016-12/documents/mdl-procedure\_rev2\_12-13-2016.pdf"
}

@article{Ye:2021jkp,
    author = "Ye, Jiaxuan and others",
    title = "{Development of water extraction system for liquid scintillator purification of JUNO}",
    eprint = "2109.07317",
    archivePrefix = "arXiv",
    primaryClass = "physics.ins-det",
    doi = "10.1016/j.nima.2021.166251",
    journal = "Nucl. Instrum. Meth. A",
    volume = "1027",
    pages = "166251",
    year = "2022"
}

@article{Landini:2024ohd,
    author = "Landini, C. and others",
    title = "{Distillation and gas stripping purification plants for the JUNO liquid scintillator}",
    eprint = "2406.01381",
    archivePrefix = "arXiv",
    primaryClass = "physics.ins-det",
    doi = "10.1016/j.nima.2024.169887",
    journal = "Nucl. Instrum. Meth. A",
    volume = "1069",
    pages = "169887",
    year = "2024"
}

@article{diVacri:2020aqc,
    author = "di Vacri, M. L. and Arnquist, I. J. and Scorza, S. and Hoppe, E. W. and Hall, Jeter",
    title = "{Direct method for the quantitative analysis of surface contamination on ultra-low background materials from exposure to dust}",
    eprint = "2006.12746",
    archivePrefix = "arXiv",
    primaryClass = "physics.ins-det",
    doi = "10.1016/j.nima.2021.165051",
    journal = "Nucl. Instrum. Meth. A",
    volume = "994",
    pages = "165051",
    year = "2021"
}

@article{JUNO:2021kxb,
    author = "Abusleme, Angel and others",
    collaboration = "JUNO",
    title = "{Radioactivity control strategy for the JUNO detector}",
    eprint = "2107.03669",
    archivePrefix = "arXiv",
    primaryClass = "physics.ins-det",
    doi = "10.1007/JHEP11(2021)102",
    journal = "JHEP",
    volume = "11",
    pages = "102",
    year = "2021"
}

@article{Li:2024gqe,
    author = "Li, Yuanxia and Zhao, Jie and Ding, Yayun and Hu, Tao and Ye, Jiaxuan and Fang, Jian and Wen, Liangjian",
    title = "{A practical approach of measuring 238U and 232Th in liquid scintillator to sub-ppq level using ICP-MS}",
    eprint = "2405.06326",
    archivePrefix = "arXiv",
    primaryClass = "physics.ins-det",
    doi = "10.1016/j.radphyschem.2025.112579",
    journal = "Radiat. Phys. Chem.",
    volume = "230",
    pages = "112579",
    year = "2025"
}

@misc{film,
    howpublished = "Daio Paper Corporation, https://www.daio-paper.co.jp/en/product/function/"
}

@article{Li:2023rae,
    author = "Li, Yuanxia and others",
    title = "{Study on U/Th residual radioactivity in acrylic from surface treatment}",
    eprint = "2301.04902",
    archivePrefix = "arXiv",
    primaryClass = "physics.ins-det",
    doi = "10.1088/1748-0221/18/05/P05023",
    journal = "JINST",
    volume = "18",
    number = "05",
    pages = "P05023",
    year = "2023"
}

@misc{IEC61709:2017,
  title        = {{IEC 61709 Electronic components - Reliability - Reference conditions for failure rates and stress models for conversion}},
  author       = {{International Electrotechnical Commission}},
  howpublished = {International Standard},
  year         = {2017},
  note         = {Third edition},
  url          = {https://webstore.iec.ch/en/publication/28554},
}

\end{document}